\documentclass{article}
\usepackage{amsmath,graphicx}

\usepackage{enumitem}
\usepackage{url}
\usepackage{graphicx}
\usepackage{float}
\usepackage{amsfonts}
\usepackage{multirow}
\usepackage{color}
\setlist{nosep, leftmargin=14pt}
\usepackage{authblk}
\usepackage{geometry}
\begin{document}


\title{When are Diffusion Priors Helpful in Sparse Reconstruction? \\ A Study with Sparse-view CT}
%
\author[1,3,$\dagger$]{Matt Y. Cheung}
\author[2,3,$\dagger$]{Sophia Zorek}
\author[3]{Tucker J. Netherton}
\author[3]{Laurence E. Court}
\author[4]{Sadeer Al-Kindi}
\author[1]{Ashok Veeraraghavan}
\author[1]{Guha Balakrishnan}
\affil[1]{Department of Electrical $\&$ Computer Engineering, Rice University, Houston TX}
\affil[2]{Department of Statistics, Rice University, Houston TX}
\affil[3]{Department of Radiation Physics, The University of Texas MD Anderson Cancer Center, Houston TX}
\affil[4]{Department of Cardiology, DeBakey Heart and Vascular Center, Houston TX}

\maketitle
\begin{abstract}
Diffusion models demonstrate state-of-the-art performance on image generation, and are gaining traction for sparse medical image reconstruction tasks. However, compared to classical reconstruction algorithms relying on simple analytical priors, diffusion models have the dangerous property of producing realistic looking results \emph{even when incorrect}, particularly with few observations. We investigate the utility of diffusion models as priors for image reconstruction by varying the number of observations and comparing their performance to classical priors (sparse and Tikhonov regularization) using pixel-based, structural, and downstream metrics.
We make comparisons on low-dose chest wall computed tomography (CT) for fat mass quantification.
First, we find that classical priors are superior to diffusion priors when the number of projections is ``sufficient''.
Second, we find that diffusion priors can capture a large amount of detail with very few observations, significantly outperforming classical priors. 
However, they fall short of capturing all details, even with many observations.
Finally, we find that the performance of diffusion priors plateau after extremely few ($\approx$10-15) projections.
Ultimately, our work highlights potential issues with diffusion-based sparse reconstruction and underscores the importance of further investigation, particularly in high-stakes clinical settings.
\end{abstract}

\renewcommand{\thefootnote}{} 
\footnotetext{$\dagger$: Equal contribution. M.C. would like to acknowledge support from a fellowship from the Gulf Coast Consortia on the NLM Training Program in Biomedical Informatics and Data Science T15LM007093. S.Z. would like to acknowledge that this material is based upon work supported by the U.S. Department of Energy, Office of Science, Office of Advanced Scientific Computing Research under Award Number DE-SC0025528. T.N. would like to acknowledge the support of the NIH LRP award.}
\renewcommand{\thefootnote}{\arabic{footnote}}

\section{Introduction}
\label{sec:intro}

\begin{figure*}[!t]
    \centering
    \includegraphics[width=\textwidth]{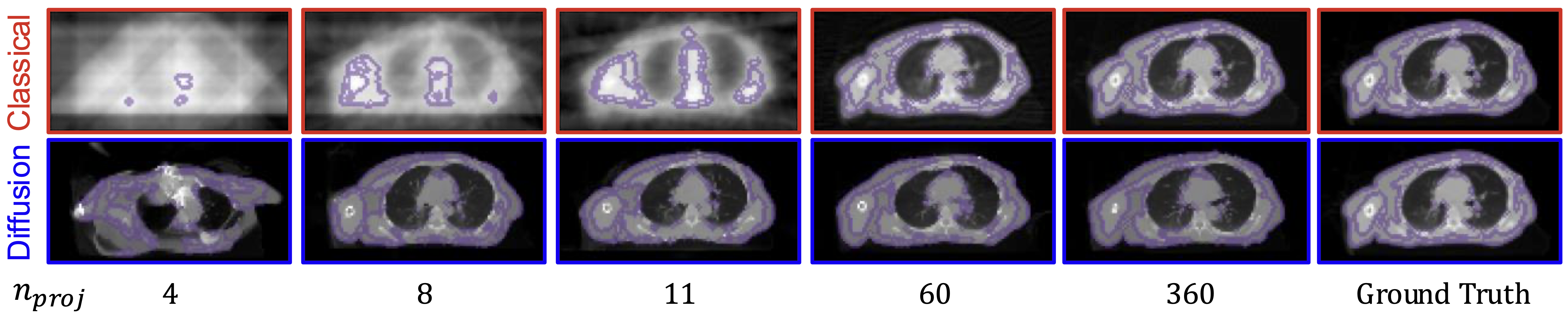}
    \caption{\textbf{Diffusion priors produce convincing results with extremely few projections. }We show reconstructions using classical (top) and diffusion (bottom) priors for varying number of projections $n_{proj}$. We show the fat segmentation in purple. Under limited $n_{proj}$, we observe the diffusion method perceptually outperforming the classical priors, but under sufficiently large $n_{proj}$, classical priors produce reconstructions whose fat segmentation's adhere more closely to the ground truth.}
    \label{fig:halluci-fat}
\end{figure*}

Medical image modalities, such as magnetic resonance imaging (MRI) and computed tomography (CT) scans, are typically reconstructed from a set of projections. For example, CT algorithms combine many x-ray projections to produce an image volume. These volumes may then be used on various downstream tasks, such as disease diagnosis and fat measurement. Image reconstruction accuracy and reliability can, therefore, directly impact the patient's standard of care.

Generally, the more projections used in reconstruction, the better the image quality. However, in settings requiring limited resources~\cite{court2023addressing} or radiation dose~\cite{rampinelli2012low}, the number of projections can be small ($<100$ instead of $100$s).
In this scenario, image reconstruction problems are ill-posed - meaning the number of projections is insufficient to fully capture the underlying anatomy of the patient.
Many past works have addressed this ``ill-posedness'' through different priors, which drive the algorithm towards a solution with certain desirable characteristics~\cite{sun2024difr3ct}.
For example, an $L_1$ prior drives the algorithm towards a solution that is sparse in some domain ~\cite{zhang2018regularization}.
Another example is $L_2$, or Tikhonov, prior that drives the algorithm towards a solution that is smooth~\cite{zhang2018regularization}.

Unlike classical priors based on simple analytical forms, an emerging alternative is to use deep generative models trained on large data distributions with neural networks. The most powerful current deep generative family are \emph{diffusion models}~\cite{dhariwal2021diffusion}, which demonstrate impressive results on a range of image generation tasks. Several recent studies show that diffusion models can produce convincing results in sparse image reconstruction with very limited projections~\cite{sun2024difr3ct} where classical priors struggle. However, diffusion models are not yet safe for deployment in practice because they can generate reconstructions that superficially look correct, but contain fabricated anatomical detail. This discrepancy highlights the need for a more nuanced evaluation framework. 

While PSNR and SSIM provide important quantitative evaluation for a machine learning practitioner, they do not link image reconstruction with downstream clinical applications and are, therefore, not meaningful and practical~\cite{cheung2024metric}.
Since prior work does not fully cover the entire range of projections or assess downstream metrics comprehensively, key questions emerge: 1) Do diffusion priors outperform classical priors across a wide range of projections (e.g., hundreds of X-rays for CT scans)? and 2) Do downstream metrics align with trends observed in pixel- and structure-based metrics?
We tackle these questions by comparing classical ($L_1$ and $L_2$) and diffusion priors over a wide range of projections and find interesting results.

We explore the real-life, practical, and clinically relevant problem of quantifying thoracic fat mass and its distribution using low-dose CT. Quantifying thoracic fat mass and its distribution is crucial to understanding its role in cardiometabolic risk and other health outcomes~\cite{wang2014imaging,dey2012epicardial}.
Traditional methods, such as full-projection computed tomography (CT) scans, offer high accuracy but are associated with significant levels of ionizing radiation. 
This concern is particularly important for individuals with obesity, who often require higher radiation doses due to increased tissue density~\cite{yanch2009increased}. 
Reducing radiation exposure while maintaining diagnostic accuracy is, therefore, critical for population-level screening and individual-level longitudinal assessments.
Using a low number of projections (low-dose CT) offers a promising solution to balance radiation safety and diagnostic effectiveness. 

Our key findings are as follows: 1) diffusion priors excel in extremely sparse settings often succeeding only with several projections, 2) classical priors are superior to diffusion priors when the number of projections is ``sufficient'', and 3) diffusion prior performance plateaus after few ($\approx$10-15) projections.
Ultimately, our work paves the way for more meaningful, practical, and clinically relevant comparisons between medical image reconstruction methods in sparse regimes.
\section{Experimental Setup}
\textbf{Dataset. } We use a de-identified CT dataset of 935 patients from The University of Texas MD Anderson Cancer Center.
All CT images were of patients who had received surgical mastectomy to the right side of the body, and radiotherapy to the post-mastectomy chest wall and/or axillary lymph nodes. We resample each scan to 1 $mm^3$, remove table artifacts and resize each slice to $128^2$ pixels. We split our dataset into 729 patients consisting of over 300k slices for training and 206 patients consisting of over 618 slices testing. We evaluate our models on test data only.

\noindent\textbf{Unconditional Diffusion Model Training. } We train a classic 2D U-Net diffusion model to reconstruct CT slices unconditionally. Our U-Net consists of four downsampling and upsampling modules with attention heads in the third and fourth layers. We use 1000 training timesteps and a squared cosine noise scheduler as outlined in \cite{ho2020denoising}. 
Model choice was determined based on loss convergence.

\begin{figure*}[!t]
    \centering
    \includegraphics[width=\linewidth]{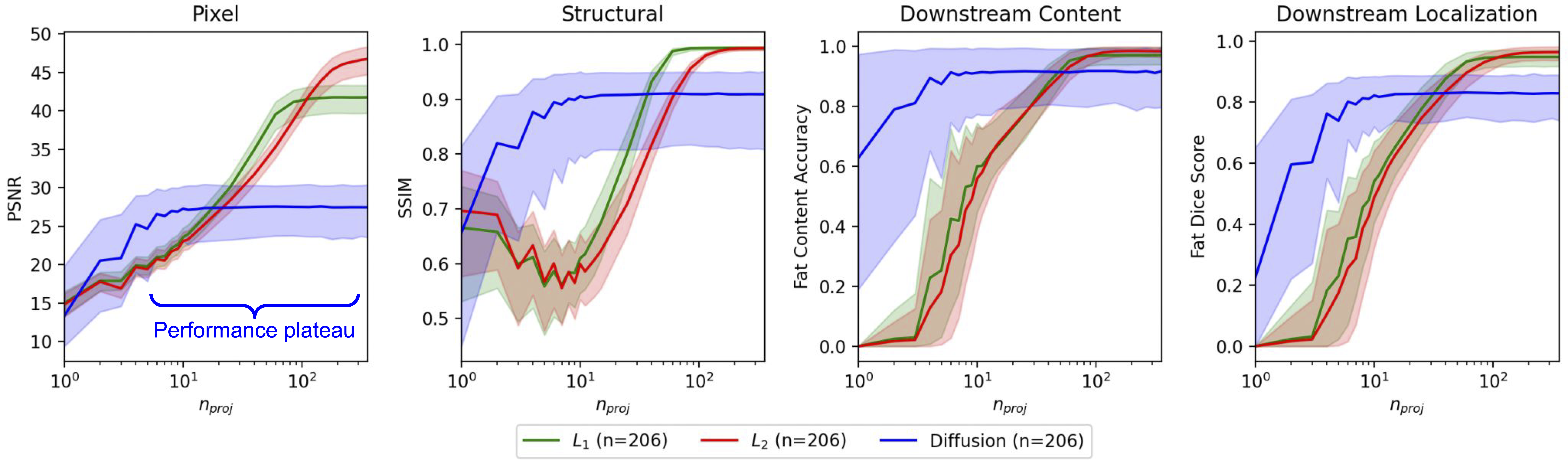}
    \caption{\textbf{Diffusion priors prevail with extremely few projections but may have wrong content and structure. } We plot evaluation metric versus number of projections for classical and diffusion priors. For 206 patients (3 slices/patient), we plot the metric and interquantile range (0.05 and 0.95 quantile, $IQR_{5,95}$) for PSNR, SSIM, fat content accuracy, and fat dice score.}
    \label{fig:trend}
\end{figure*}

\noindent\textbf{Inference. }
We performed a grid search to determine the best optimization, regularization, and guidance parameters for $L_1$ ($\lambda=10$), $L_2$ ($\lambda=10$) and diffusion ($\lambda=0.1$).
These corresponded to the lowest mean squared error at 360 projections 
We used the \verb|tomosipo| package~\cite{hendriksen2021tomosipo} to compute the projection operator.
For $L_1$, we ran the total-variation regularized least squares for reconstruction for 1000 iterations. 
At each time step of our guided diffusion inference we take the projection of the predicted reconstruction using our projection operator and compute its mean squared error with the ground truth projections, varying the number of projections used. We then compute the gradients of this loss with respect to the predicted reconstruction, and guide the diffusion process by adding the gradients back to the prediction~\cite{dhariwal2021diffusion,graikos2022diffusion}. We iterate through this process over 50 timesteps.

\noindent\textbf{Metrics}
We investigate pixel (peak-signal-to-noise ratio or PSNR), structural (structural similarity index or SSIM), and downstream metrics (content and localization).
We use the percentage of correct pixels classified as fat (fat content accuracy) and the fat segmentation dice score (fat localization accuracy).
Fat has Hounsfield units between -150 and -50.
\section{Results}

\noindent\textbf{Diffusion priors work well with extremely few projections. }
We find that diffusion priors excel in extremely sparse settings to estimate pixel-based, structural-based, and downstream metrics, often already succeeding in capturing many details only with several projections (Fig. \ref{fig:halluci-fat}).
However, they fall short of capturing all details accurately, like vasculature, even with many observations.
On the other hand, we find that classical priors are superior to diffusion priors when the number of projections is ``sufficient'' (i.e., exceeds extremely few projections).
We leverage results from Fig. \ref{fig:trend} to identify the regions where diffusion priors are \textit{guaranteed} to be better than classical priors. 
We can define the number of projections ($n_{proj}$) as the interval $I=[LB,UB]$ where diffusion priors perform statistically better than classical priors.
The lower and upper bounds are denoted as $LB,UB\in\mathbb{R}\times\mathbb{R}$.
For a monotonically increasing (can be positive or zero slope) metric ($M$) as number of projections ($n_{proj}$) increases, the lower bound is given by $n$ at the minimum acceptable metric ($\tau_M$), and the upper bound is given by the maximum $n_{proj}$ when the Mann-Whitney U-test has a p-value of less than a user-defined threshold ($\tau_p$; for example, $0.05$): $I=[\min \{n_{proj}\in\mathbb{R}: M\geq\tau_M\}, \max\{n_{proj}\in\mathbb{R}:p\leq\tau_p\}]$.
The upper bound of the interval gives the highest number of projections where the probability that a randomly chosen metric value generated from one method is larger than a randomly chosen metric value generated from another $P(M_1>M_2)$.
For a monotonically decreasing $M$, maximum and minimum are used for the lower and upper bounds.
We choose the lower bound to be the ``corner'' of the plots where performance starts to plateau.
We show these bounds in Table. \ref{tab:bounds}.
We find that diffusion priors perform better than classical priors between extremely few observations (5 to 10) and 10s of projections.
Beyond $10$s of projections, classical priors perform better overall.
\begin{table}[t]
\label{tab:bounds}
\centering
{%
\begin{tabular}{c|cl|cc|}
\cline{2-5}
 & \multicolumn{2}{c|}{\multirow{2}{*}{LB}} & \multicolumn{2}{c|}{UB (p-value)} \\ \cline{1-1} \cline{4-5} 
\multicolumn{1}{|c|}{\textbf{Metric}} & \multicolumn{2}{c|}{} & \multicolumn{1}{c|}{$L_1$ Prior} & $L_2$ Prior \\ \hline
\multicolumn{1}{|c|}{MSE} & \multicolumn{2}{c|}{7} & \multicolumn{1}{c|}{15 (3.7e-40)} & 15 (2.4e-42) \\ \hline
\multicolumn{1}{|c|}{SSIM} & \multicolumn{2}{c|}{7} & \multicolumn{1}{c|}{25 (4.24e-120)} & 60 (7.4e-40) \\ \hline
\multicolumn{1}{|c|}{Fat Content} & \multicolumn{2}{c|}{9} & \multicolumn{1}{c|}{40 (9.1e-43)} & 40 (1.25e-72) \\ \hline
\multicolumn{1}{|c|}{Fat Localization} & \multicolumn{2}{c|}{7} & \multicolumn{1}{c|}{25 (1.2e-69)} & 25 (5.39e-120) \\ \hline
\end{tabular}%
}
\caption{\textbf{Diffusion priors perform statistically better than classical priors for extremely low number of projections.} Upper (UB) and Lower (LB) bounds for number of projections where diffusion performs better than classical priors. The lower bound is from the ``corner'' of the performance curve and upper bound is based on the Mann-Whitney U-test.
}
\end{table}

\noindent\textbf{Diffusion prior performance plateaus after extremely few projections. }
We observe that as the number of projections gets larger, the performance curves plateau and do not reach perfect reconstruction (Fig. \ref{fig:trend}).
In contrast, the performance of classical priors continually increases and supersedes diffusion priors beyond extremely few projections.
Furthermore, we show the inter-quantile range for quantiles 0.05 to 0.95 ($IQR_{5,95}$) in Fig. \ref{fig:trend}.
While the diffusion prior curves are heteroscedastic for extremely few projections, they get more homoscedastic when the performance plateaus.
These observations show that the diffusion priors do not leverage the extra information well for reconstruction beyond extremely few projections.
Thus, diffusion priors get high-level structures (i.e., the shape of the body) correct but not low-level structures (i.e., vasculature) (Fig. \ref{fig:halluci-fat}).

\section{Conclusion and Discussion}
In this paper, we answer the question ``when are diffusion priors helpful for sparse reconstruction?'' by comparing classical priors used in practice with state-of-the-art diffusion priors for different number of projections and metrics.
First, we find that classical priors are superior to diffusion priors when the number of projections is ``sufficient''.
Second, we find that diffusion priors can capture a large amount of detail with very few observations, significantly outperforming classical priors. 
However, they fall short of capturing all details, even with many observations.
Finally, we find that the performance of downstream metrics plateau after few ($\approx$10-15) projections for diffusion priors.

One possible explanation for the performance saturation of diffusion priors is guided diffusion with naive gradient descent may not converge to the optimal solution.
While the unconditional model steers the gradient toward the data space, the gradient of the external loss (i.e., the mean square error between predicted projections and ground truth projections) may point away from the data space~\cite{guo2024gradient}.
The non-optimal convergence may explain why, even with more projections, $IQR_{5,95}$ remains constant because the guidance process eventually becomes ``stuck'' balancing the steering of the diffusion prior and minimizing the external loss.
Therefore, when the number of projections is high, they still fail to capture the low-level features like vasculature.
A possible solution could be adaptive fine-tuning \cite{guo2024gradient}.

\textbf{Clinical Implications. }While diffusion priors may not capture low-level details well, we demonstrate the potential of diffusion priors for accurate assessment of thoracic fat quantity.
Using extremely few projections significantly reduces the radiation exposure required for imaging, lowering the dose by up to 97\% compared with full projection and by up to 75\% compared to classical priors. 
The reduction in radiation is particularly advantageous for individuals who require repeated imaging or those with obesity, where standard imaging protocols typically involve higher radiation doses.
The enhanced performance of low-dose CT scalability makes it a viable solution for large-scale studies and clinical applications where minimizing radiation is critical. 
Fixed planar X-ray systems could be employed in lieu of expensive CT machines, making diffusion priors especially valuable in low-resource settings where access to advanced imaging technologies is limited.
\section{Compliance with Ethical Standards}
This research was conducted retrospectively using human data under an approved IRB protocol.

\bibliographystyle{IEEEbib}
\bibliography{refs}

\end{document}